%% file: Active_Cell.tex
\documentclass[aps,pre,twocolumn]{revtex4}
\usepackage{epsfig,amssymb,amsmath,amsthm,amsfonts,amsbsy,mathrsfs}
\usepackage{graphicx}
\usepackage{color}
\usepackage{tikz}
\usetikzlibrary{arrows,shapes,chains}
\newcommand*\circled[1]{\tikz[baseline=(char.base)]{
\node[shape=circle,draw,inner sep=0.5pt] (char) {#1};}}

\usepackage{pifont}

\definecolor{cream}{RGB}{222,217,201}

\begin{document}
\title{Relaxation Dynamics in Persistent Epithelial Tissues}
\author{Meng-Yuan Li}
\author{Yan-Wei Li}
\email{yanweili@bit.edu.cn}
\affiliation{Key Laboratory of Advanced Optoelectronic Quantum Architecture and Measurement (MOE), School of Physics, Beijing Institute of Technology, Beijing, 100081, China}
\date{\today}

\begin{abstract}
Cell monolayers and epithelial tissues display slow dynamics during the liquid-glass transitions, a phenomenon with direct relevance to embryogenesis, tumor metastases, and wound healing. In active cells, persistent motion and cell deformation compete, significantly influencing relaxation dynamics. Here, we numerically construct the liquid-glass transition phase diagram for two-dimensional polydisperse persistent cells. We employ cage-relative measures and conduct extensive simulations to eliminate the influence of system size effects. These effects arise from long-wavelength fluctuations in nearly equilibrated cells and a combination of long-wavelength fluctuations and non-equilibrium effects in highly persistent cells. Our study unveils distinctive intermittent dynamics associated with intermittent T1 transitions in highly persistent cells, where the velocity correlates over space with a characteristic length $\xi$. The $\alpha$ relaxation time exhibits a universal power-law dependence on the irreversible T1 transition rate, $\Gamma_{\rm{T1}}^{\rm irr}$, multiplied by ${\rm exp}(\xi)$. Here, $\xi$ vanishes in nearly equilibrated cells, and $\Gamma_{\rm{T1}}^{\rm irr}$ diminishes towards the mode-coupling glass transition point.
\end{abstract}
\maketitle

Active systems harness energy from the environment, drive themselves significantly beyond equilibrium, and induce persistent motion among their constituents~\cite{Aditi_Review, Bechinger_Review}. Among these, active cells are intensely explored~\cite{Bi_prx, Anshuman_SM,melting_activecells,Bi_prx2021,Kim_NM,Simon_pnas,Luca_NP,Barton2017}. It is found that depending on cell deformability and mobility, active cells in biological tissues exhibit a reversible transition from fluid-like mesenchymal cells to a solid-like epithelial layer~\cite{Bi_prx, Anshuman_SM,melting_activecells}. These transitions play crucial roles in various biological processes, such as organ development, tumor metastases, and wound healing, et al~\cite{Kim_NM,Levine_PNAS,Silberzan_pnas,Cornelis,Manning2013,Weitz_PNAS2011,EMT1, EMT2, EMT3}.

Recent experiments and simulations revealed that this mesenchymal to epithelial transition shares many similarities with the glass transition~\cite{Park_NM,Puliafito_PNAS,Angelini_prl,Manning_EPL,ywli_pre}. For instance, supercooled dynamics occur during this transition. These include the caging behavior where cells are confined by their neighbors, a two-step relaxation process involving initial rattling of cells within the cage ($\beta$-relaxation) succeeded by long-range motion following cage escape ($\alpha$-relaxation), and the collective and heterogeneous dynamics~\cite{Park_NM,Puliafito_PNAS,Manning_EPL,ywli_pre}. Moreover, novel intermittent relaxation events have been identified in active particles with large persistent time in two-dimensional particulate systems~\cite{Mandal_nc2020,Berthier_prl2022,Berthier_sm2023}, wherein the system relaxes through sequences of jumps between mechanical equilibriums, arising from the equilibrium between self-propulsion forces and interacting forces. These findings prompt further exploration into the relaxation dynamics of active cells, where interactions are of many-body type, as commonly observed in various soft materials~\cite{dense_granular,soft_jammed,polyelectrolyte_star,microgels,Star_polymer,Likos_star,Dendrimer}, and structural rearrangements heavily rely on the topological T1 processes~\cite{Rivier_1984, Staple2010, Bi_softMatter,Bi_prx2021,ywli_pre}.

In this Letter, we explore the interplay between deformability and the persistent time in active cells and investigate the resulting relaxation dynamics. Our analysis reveals hitherto unexplored significant system size effects stemming from long-wavelength fluctuations in cells with small persistence, effectively addressed through cage-relative (CR) measures. Conversely, extended systems are required to account for additional non-equilibrium effects at large persistence. After mitigating these size effects, we construct a dynamical arrest phase diagram in the plane of persistent time and deformability. We identify intermittent kinetic energy and the associated intermittent T1 transitions at large persistence, shaping a distinctive relaxation process. The distinct relaxation modes at small and large persistent times are encapsulated in a universal function describing the close correlation of the $\alpha$ relaxation time with the irreversible T1 transition rate and the length scale of spatial velocity correlations.

\begin{figure*}[tb]
 \centering
 \includegraphics[angle=0,width=0.8\textwidth]{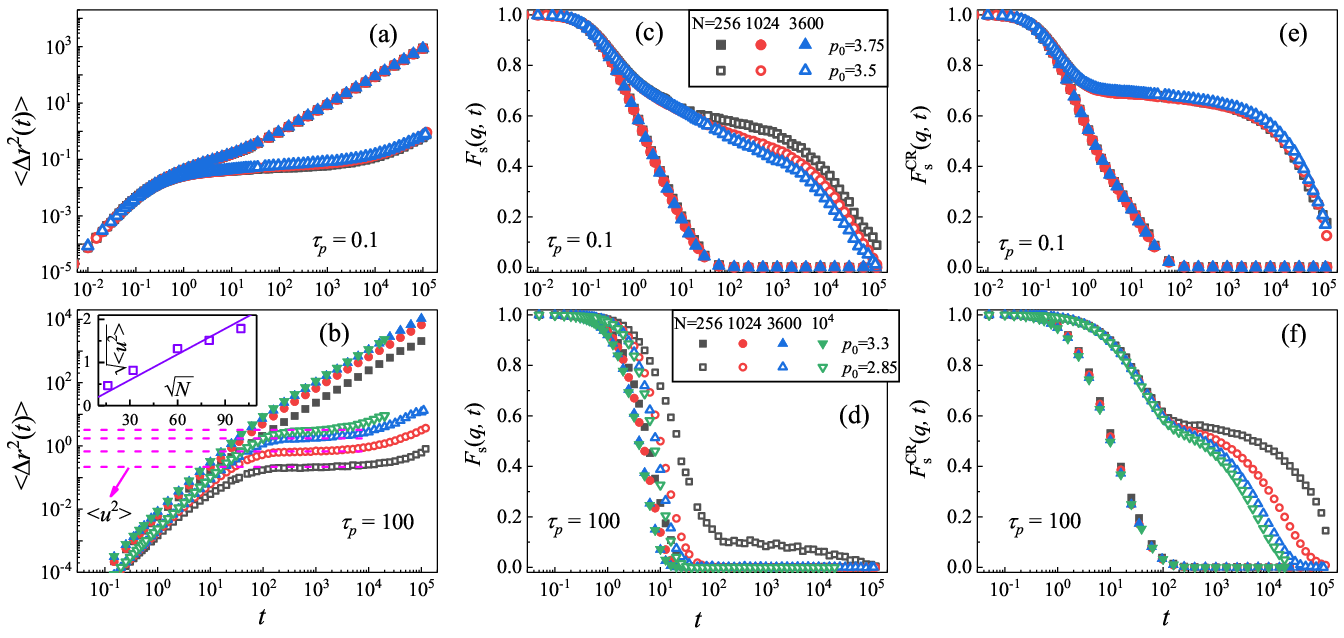}
 \caption{Time dependence of the mean square displacement (a), of the self-intermediate scattering function (c), and of the
cage-relative self-intermediate scattering function (e), at persistence time $\tau_p=0.1$. Close and open symbols are for $p_0=3.75$ and $p_0=3.5$, respectively. Panels (b), (d) and (f) show the time dependence of the same quantities for $\tau_p=100$, at $p_0=3.3$ (close symbols) and at $p_0=2.85$ (open symbols). Different colors of the symbols in all main panels indicate different system sizes ranging from $N=256$ to $N=3600$ for $\tau_p=0.1$, and is up to $10000$ for $\tau_p=100$. Magenta dashed lines in panel (b) mark the Debye-Waller factor $\langle u^2\rangle$ evaluated from the plateau of the mean square displacement, and (b) inset demonstrates a linear dependence of $\langle u^2\rangle$ on $\sqrt N$.
\label{fig:dyn}
}
\end{figure*}

We study the dynamics of the polydisperse self-propelled Voronoi (SPV) model describing an active system of confluent epithelial cells~\cite{Frank_vertex, Staple2010, Bi_np2015, Manning1,Vertex_model, Bi_prx}. The configurational degrees of freedom are the centers of mass of the cells, $\{\mathbf{r}_i\}$, and their Voronoi tessellations determine the shapes of cells. The mechanical energy of a cell depends on its area $A_i$ and perimeter $P_i$, $E_i = K_A (A_i-A_{0}^i)^{2}+K_P(P_i-P_{0}^i)^{2}$, where $A_0^i$ and $P_0^i$ are preferred values, while $K_A$ and $K_P$ are area and perimeter elastic constants. The two quadratic terms are from biological considerations~\cite{Frank_vertex,  Staple2010, Bi_np2015, Manning1,Vertex_model, Bi_prx,Angelini_prl,giavazzi2018flocking}.
Hence, the dimensionless energy functional is
\begin{equation}
e=\sum_{i=1}^{N}[(a_{i}-a_{0}^{i})^{2}+r^{-1}
(p_{i}-p_{0}^{i})^{2}],
\label{eq:E}
\end{equation}
where $a_i = A_i/l^2$ and $p_i = P_i/l$ with $l$ the unit of length which we have chosen so that $\langle a_{i} \rangle=1$. We fix the inverse perimeter modulus, $r = K_A l^2/K_P$ to unity. The preferred area $a_{0}^{i}$ is uniformly distributed in the range $0.8$--$1.2$, to avoid crystallization, while the preferred perimeter is fixed to $p_{0}^{i}=p_0\sqrt{a_{0}^{i}}$, with $p_0$ the target shape index, which determines the cell deformability and higher values of $p_0$ corresponding to more deformable cells~\cite{Bi_np2015, Ourwork_PRM}.

Cell centers $\{\mathbf{r}_i\}$ evolve according to the overdamped equations of motion $\zeta\frac{d\mathbf{r}_i}{dt} = {\bf f}_i^m + {\bf f}_i^a $~\cite{Bi_prx}, where ${\bf f}_i^m = -{\bf{\nabla}}_ie$ is the mechanical interaction force that  is of many-body type as it cannot be expressed as a sum of pairwise forces. $\zeta$ is the frictional coefficient, which also sets our time unit $\tau=\zeta/(K_Al^2)$. ${\bf f}_i^a =v_0\hat{n}_i$ is the self-propulsion force with constant magnitude $v_0=1$ and directed along the polarity vector $\hat{n}_i = (\cos\theta_i,\sin\theta_i)$, where the polarity angle $\theta_i$ is perturbed by $\partial_t\theta_i = \eta_i$  with $\eta_i$ a Gaussian white noise of zero mean and variance $\langle\eta_i(t)\eta_j(t')\rangle = 2D_r\delta(t-t')\delta_{ij}$. Here, $D_r$ is the rotational diffusion coefficient resulting in the persistent time $\tau_p=1/D_r$. When $\tau_p\rightarrow0$, equilibrium Brownian motion is recovered. In the following, we investigate the relaxation as the shape index $p_0$ and the persistent time $\tau_p$ are varied.

We carry out simulations in a square box under periodic boundary conditions~\cite{Allen_book}. The number density is set to be 1. Voronoi tessellations are computed using the Boost C++ Voronoi library~\cite{boost}. The number of cells $N$ ranges from $256$ to $10^4$, and $N=3600$ for most of our results, if not otherwise specified. We discuss below the important effect of system size on the relaxation dynamics of the active cells. 

A rigidity transition has been identified at zero temperature by decreasing the target shape index $p_0$ to $p_0 \simeq 3.81$~\cite{Bi_np2015, Ourwork_PRM}. In active cells, a similar dynamical arrest transition occurs as $p_0$ decreases, with the transition point contingent on the values of the persistent time $\tau_p$~\cite{Bi_prx}. We first characterize the $\tau_p$-dependent liquid-glass transition boundary by investigating the relaxation dynamics. This analysis considers the previously unexplored system size effect, and we further delve into its origin. We study the mean square displacement (MSD) $\left\langle\Delta r^{2}(t)\right\rangle=\left\langle\frac{1}{N} \sum_{i=1}^{N} \Delta \mathbf{r}_{i}(t)^{2}\right\rangle$, where $\Delta \mathbf{r}_{i}(t)$ represents the displacement of particle $i$ during time $t$, and the self-intermediate scattering function (ISF) $F_{s}(q, t)=\left\langle\frac{1}{N} \sum_{j=1}^{N} e^{i\mathbf{q} \cdot \Delta \mathbf{r}_{j}(t)}\right\rangle$ with $q=|\mathbf{q}|$ the wavenumber of the first peak of the static structure factor. The MSD exhibits ballistic behavior at short times, $\left\langle\Delta r^{2}(t)\right\rangle\sim v^2t^2$, with $v^2=\langle \mathbf{v}_i^2\rangle$ being the mean square velocity. We observe a convergence of MSD in the ballistic regime at $\tau_p=0.1$, while $v^2$ decreases as $p_0$ is reduced at $\tau_p=100$, suggesting a significant impact of activity on the velocities, as discussed in detail in Fig. S2 in the Supplemental Material (SM)~\cite{sm}. This observation is analogous to that in the dense assembly of persistent particles~\cite{Berthier_prl2022}. At longer times, the system directly enters the diffusive regime at large $p_0$ in the liquid states. Conversely, we find caging, characterized by the long plateau in the MSD (Figs.~\ref{fig:dyn}(a) and \ref{fig:dyn}(b)), and the emerging shoulder in the ISF (Fig.~\ref{fig:dyn}(c) and Fig.~\ref{fig:dyn}(d) for a small system size), before the system becomes diffusive at small $p_0$ in the supercooled states.

\begin{figure}[tb]
 \centering
 \includegraphics[angle=0,width=0.4\textwidth]{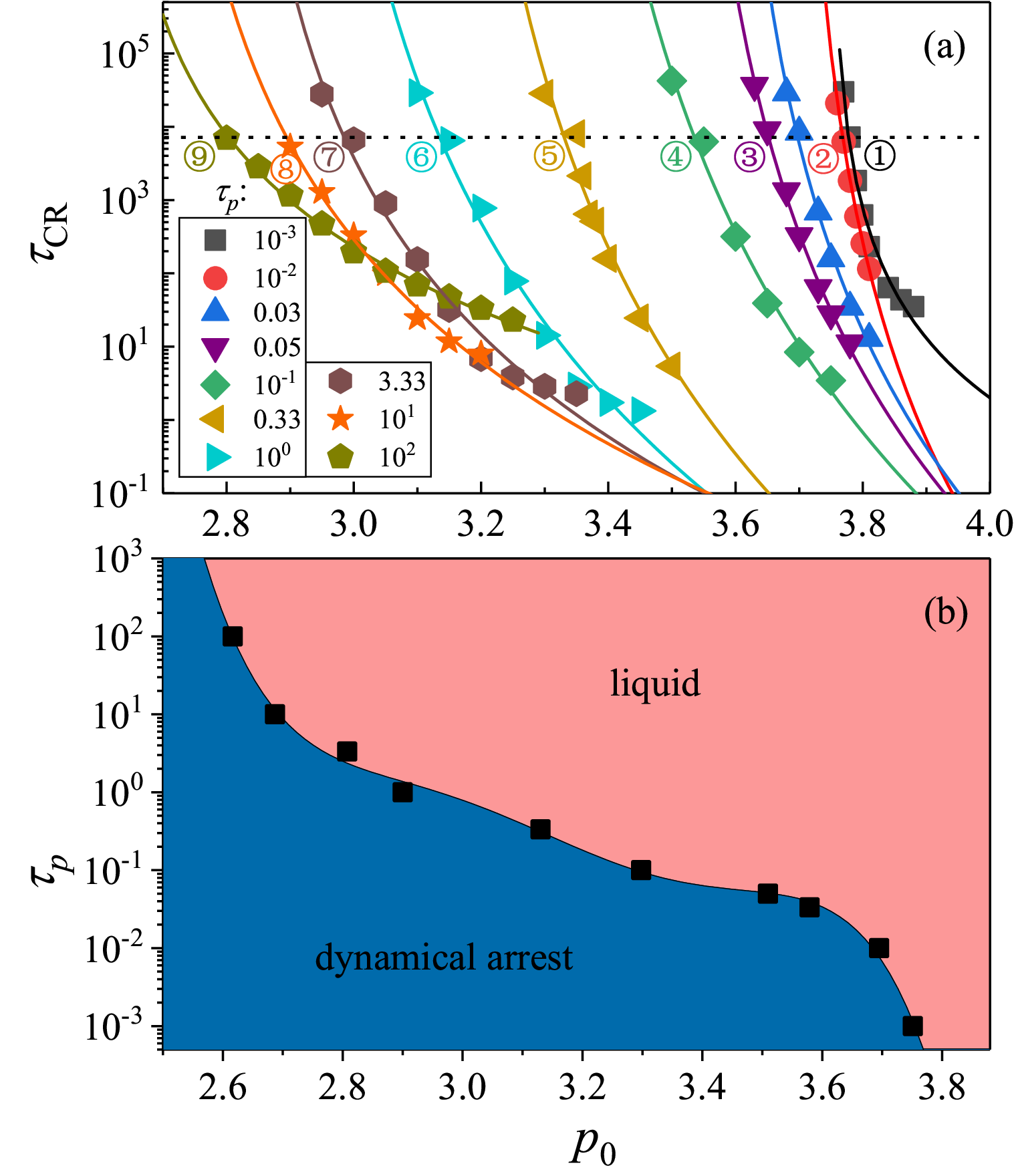}
 \caption{ 
 (a) Dependence of the cage-relative $\alpha$-relaxation time ($\tau_{\rm CR}$) on the target shape index ($p_0$) for various values of the persistent time ($\tau_p$). Solid lines represent the fitting results from the mode-coupling theory. We specifically select $9$ state points with comparable $\tau_{\rm CR}\simeq7100$, as indicated by the horizontal black dashed line. (b) The glass transition phase diagram of active cells in the $\tau_p-p_0$ plane. Black squares denote the mode-coupling glass transition point, while the boundary line is from a fifth-order polynomial fit.
\label{fig:pd}
}
\end{figure}

\begin{figure}[tb]
 \centering
 \includegraphics[angle=0,width=0.48\textwidth]{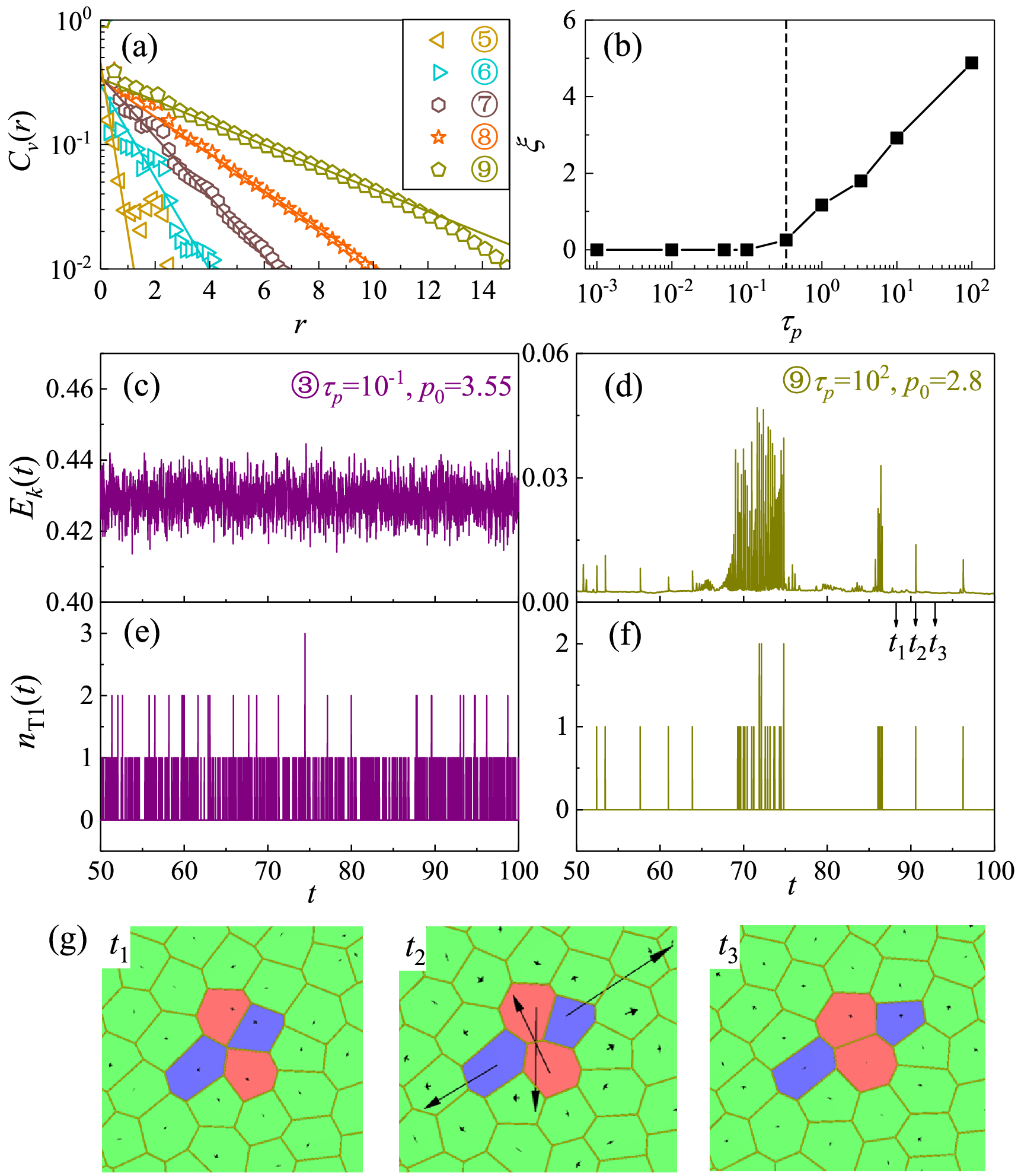}
 \caption{
(a) Spatial velocity correlation function, $c_v(r)$, for state points \ding{196}-\ding{200}. (b) The velocity correlation length $\xi$ as a function of the persistence time $\tau_p$ for the $9$ selected state points, as indicated in Fig.~\ref{fig:pd}(a). The critical value $\tau_p^c$, at which $\xi$ diverges from $0$, is denoted by the vertical dashed line in (b).
Time series of ((c) and (d)) the kinetic energy and ((e) and (f)) the instantaneous number of T1 transition for state points \ding{194} and \ding{200}. (g) Visualization of local configurations before, during, and after the T1 transition, at time $t_1$, $t_2$, and $t_3$ as indicated at the bottom of (d), for state point \ding{200}. Cells involved in the T1 transition are highlighted in blue and red, while others are green. Velocity vectors are represented by black arrows in (g). 
\label{fig:T1}
}
\end{figure}
The dynamics display a pronounced dependence on system size, especially in the supercooled regime, as illustrated by the ISF for $\tau_p=0.1$ (Fig.~\ref{fig:dyn}(c)), as well as in both the MSD and the ISF for $\tau_p=100$ (Figs.~\ref{fig:dyn}(b) and \ref{fig:dyn}(d)). One contribution of this system size dependence is the long-wavelength fluctuations, known to be relevant in two dimensions~\cite{MWPrl,Shiba,Weeks_longwave,Keim_MW,ywli_PNAS}, evidenced by the linear dependence of the square root of the Debye-Waller factor, $\sqrt{\langle u^2\rangle}$ on $\sqrt{N}$ (Fig.~\ref{fig:dyn}(b) inset). Filtering out the impact of long-wavelength fluctuations is achieved through CR measures~\cite{Shiba,Weeks_longwave,Keim_MW,ywli_PNAS}, wherein the standard displacement $\Delta \mathbf{r}_i(t)$ in dynamical quantities is replaced by the CR displacement $\Delta \mathbf{r}{_i}^{\rm CR}(t) = \Delta \mathbf{r}_i(t) - 1/N{_i}\sum_{j=1}^{N_{i}} \Delta \mathbf{r}_j(t)$, with the sum extending over the $N{_i}$ neighbors particle $i$ has at time $0$. A data collapse of the CR-ISF, $F_{s}^{\rm CR}(q, t)$, is observed for various system sizes at $\tau_p=0.1$ (Fig.~\ref{fig:dyn}(e)), where non-equilibrium effects are not prominent. At $\tau_p=100$, conversely, data collapse is observed for CR-ISF only in the short-time ballistic and caging regimes, not in the long-time $\alpha$ relaxation regime, although there is a qualitative change in the relaxation behavior from an anomalous one-step decay in the ISF (Fig.~\ref{fig:dyn}(d)) to a two-step decay in the CR-ISF (Fig.~\ref{fig:dyn}(f)). Hence, nonequilibrium effects may also contribute to the system size dependence. This necessitates investigation in a relatively large system at large persistence times. In Fig. S1 of the SM~\cite{sm}, we observe a result analogous to CR-ISF investigating the bond-orientational correlation function, which, by definition, is unaffected by long-wavelength fluctuations. In the following, we employ CR measures for quantities influenced by long-wavelength fluctuations and focus on a system with $N=3600$ for all values of $\tau_p$. This choice ensures that system size effects are negligible, even for the largest $\tau_p$ ($\tau_p=100$) studied (Fig.~\ref{fig:dyn}(f)).

\begin{figure}[tb]
 \centering
 \includegraphics[angle=0,width=0.45\textwidth]{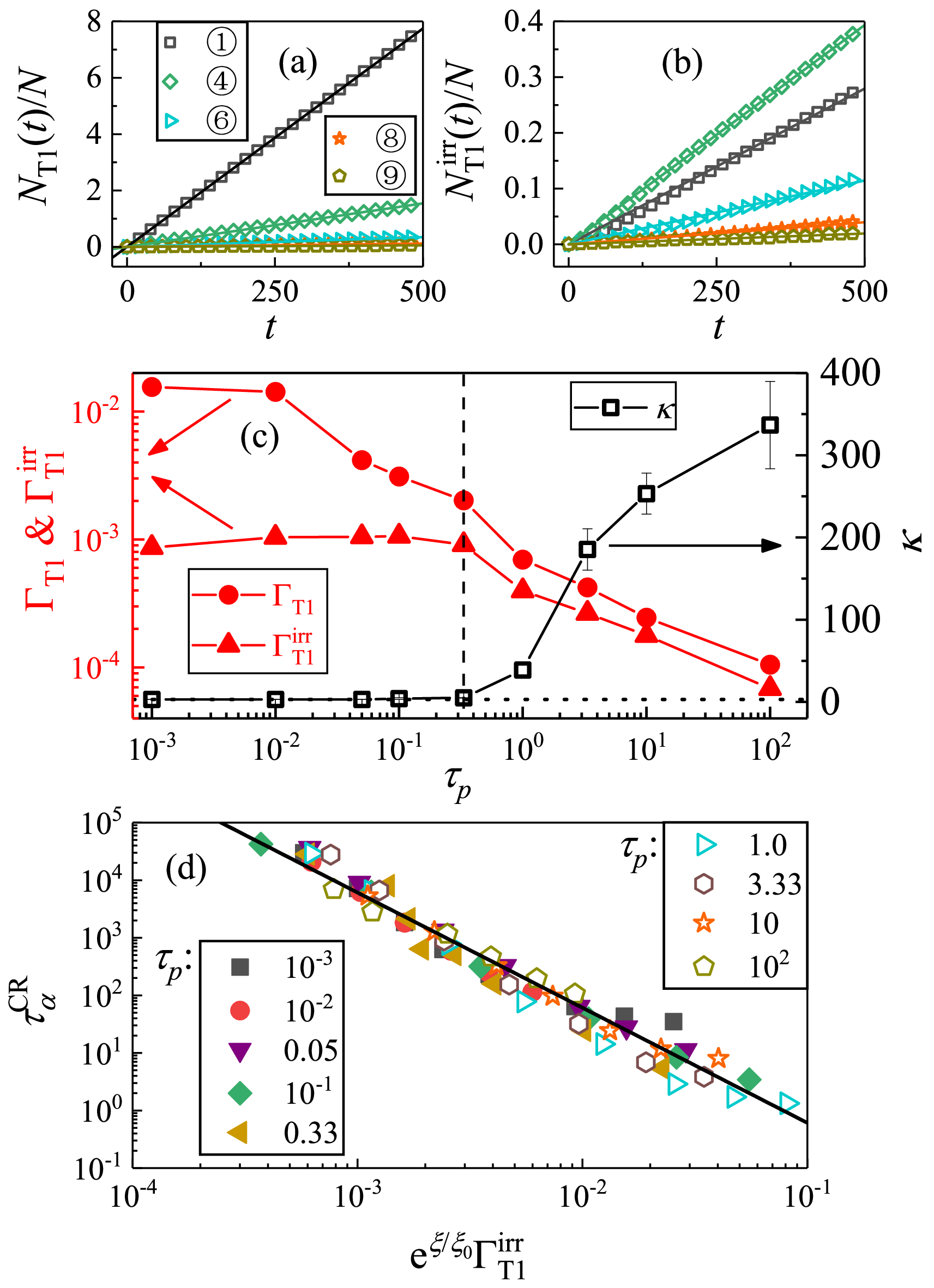}
 \caption{Time dependence of the cumulative count of (a) all T1 transitions per cell, denoted as $N_{\rm T1}/N$, and (b) irreversible T1 transitions per cell, denoted as $N_{\rm T1}^{\rm irr}/N$. (c) the kurtosis of the kinetic energy, $\kappa$ (black squares) and the rate of all ($\Gamma_{\rm T1}$) and irreversible ($\Gamma_{\rm T1}^{\rm irr}$) T1 transitions, as functions of the persistent time $\tau_p$. (d) illustrates the relaxation time $\tau_{\rm CR}$ as a function of ${\rm exp(\xi/\xi_0)}\Gamma_{\rm T1}^{\rm irr}$ at different persistent times, as detailed in the legend.
\label{fig:scaling}
}
\end{figure}

We define the CR $\alpha$-relaxation time $\tau_{\rm CR}$ as $F_{s}^{\rm CR}(q, \tau_{\rm CR})=e^{-1}$, and plot in Fig.~\ref{fig:pd}(a) $p_0$ dependence of $\tau_{\rm CR}$ for various values of $\tau_p$. Fitting $\tau_{\rm CR}(p_0)$ with the algebraic prediction of the mode-coupling theory~\cite{Ediger}, given by $\tau_{\rm CR}\sim(p_0-p_0^c)^\gamma$, enables the determination of the liquid-glass transition boundary $p_0^c$ for different $\tau_p$. $p_0^c$ is indicated by black squares in the glass transition phase diagram of the active cells, as illustrated in Fig.~\ref{fig:pd}(b). The phase diagram suggests that both deformability and persistence time impact the glass transition boundary, and $p_0^c$ increases monotonically as $\tau_p$ decreases. However, we demonstrate in the following that the relaxation dynamics differ significantly at large and small values of $\tau_p$. To this end, we select $9$ state points with comparable $\tau_{\rm CR}$ at different values of both $\tau_p$ and $p_0$, as indicated in Fig.~\ref{fig:pd}(a) (\circled{1}-\circled{9}). We first find that the persistent propulsion induces spatial correlation of the velocity, quantified by $C_v(r)=\langle \mathbf{v}(r)\cdot \mathbf{v}(0)\rangle/\langle v^2\rangle$.  In Fig.~\ref{fig:T1}(a), we illustrate $C_v(r)$ for state points \circled{1}-\circled{9}, where spatial correlation of velocity is evident. The correlation extends over a larger length scale with increasing the persistence time. We quantify the correlation length $\xi$ by fitting $C_v(r)$ with the exponential form $C_v(r)\propto \exp(-r/\xi)$. In Fig.~\ref{fig:T1}(b), we present the resulting $\tau_p$ dependence of $\xi$. Remarkably, $\xi$ acquires finite values at $\tau_p^c\simeq0.33$ and grows with increasing $\tau_p$. The increasing length scale implies the requirement for a large system size to accommodate the cells and their velocity correlation, likely serving as the origin of the observed system size effects in highly persistent cells.

In Figs.~\ref{fig:T1}(c) and \ref{fig:T1}(d), we compare the time series of kinetic energy, $E_k(t)$, for state points \circled{3} and \circled{9}, respectively, at relatively small and large values of $\tau_p$. We observe Gaussian-like fluctuations of $E_k(t)$ at small $\tau_p$, resembling observations in thermally equilibrated systems. Here, relaxation events are thermally activated, exhibiting broad distributions of energy barriers in the energy landscape. In contrast, at large $\tau_p$, $E_k(t)$ displays sudden bursts and periods of quiescence, indicating intermittent dynamics~\cite{Mandal_nc2020,Berthier_prl2022}. This behavior arises from the alternation between mechanically equilibrium states achieved through the balance between active force and interaction force, and rearrangement that drives the system from one mechanical equilibrium to another.

Cell rearrangement relies on the T1 topological transition~\cite{Rivier_1984, Staple2010, Bi_softMatter,Bi_prx2021,ywli_pre}, illustrated in Fig.~\ref{fig:T1}(g), where an edge connecting neighboring cells (in blue) disappears, and a new edge connecting previously separated cells (in red) emerges. The time series of the instantaneous number of T1 events, $n_{\rm T1}(t)$, is presented in Figs.~\ref{fig:T1}(e) and \ref{fig:T1}(f) for state points \circled{3} and \circled{9}, respectively. This count is obtained by comparing configurations at time $t$ and the previous time step. A close correspondence between $E_k(t)$ and $n_{\rm T1}(t)$ is observed. At small persistence times, $n_{\rm T1}(t)$ deviates from $0$ at random times, resembling white noise. Conversely, at large persistence times, it spikes when $E_k(t)$ surges and remains at $0$ during periods of quiescence in $E_k(t)$. This close correlation is further visualized in Fig.~\ref{fig:T1}(e), where we illustrate the local configuration at time $t_2$, precisely at the peak of the kinetic energy, along with the configurations at $t_1$ and $t_3$ before and after the surge. These three time points are marked at the bottom of Fig.~\ref{fig:T1}(d). A T1 transition at $t_2$ is identified, highlighting cells (in red and blue) that participate in this T1 transition in these configurations. The T1 rearrangement is associated with large instantaneous velocities of the involved cells, as indicated by the black vectors. Facing pair cells exhibit velocities pointing in opposite directions and display four-fold features.

Notably, as seen in Figs.~\ref{fig:T1}(e) and \ref{fig:T1}(f), T1 transitions occur more frequently at small persistence times than at large ones. We therefore consistently find a higher cumulative count of T1 transitions per cell, $N_{\rm{T1}}/N$, at lower persistent times, as demonstrated in Fig.\ref{fig:scaling}(a). The linear dependence of $N_{\rm{T1}}/N$ on time allows us to define the T1 transition rate per cell $\Gamma_{\rm{T1}}$, which decreases monotonically as the persistent time increases, as shown by the red circles in Fig.\ref{fig:scaling}(c). Remind that the CR $\alpha$ relaxation times for these state points are comparable, as indicated in Fig.~\ref{fig:pd}(a). This suggests that the rate of T1 transition may not be the key parameter governing the slow $\alpha$ relaxation. We then turn to count irreversible T1 transitions, where a second consecutive T1 transition does not reverse the first one, as reverse T1 transitions recover the local configuration and do not contribute to the relaxation. The time evolution of the cumulative number of irreversible T1 transitions per cell, $N_{\rm{T1}}^{\rm irr}/N$, is presented in Fig.\ref{fig:scaling}(b), and the rate of irreversible T1 transition, $\Gamma_{\rm{T1}}^{\rm irr}$ (red triangles), is illustrated in Fig.\ref{fig:scaling}(c) for those $9$ selected state points with comparable $\tau_{\rm CR}$ (Fig.~\ref{fig:pd}(a)). Intriguingly, $\Gamma_{\rm{T1}}^{\rm irr}$ reaches a plateau for small persistent times, $\tau_p \leq \tau_p^c\simeq0.33$. This critical value, $\tau_p^c$, coincides with the emergence of spatial velocity correlation (Fig.~\ref{fig:T1}(b)) and with the onset of intermittence in the kinetic energy, quantified by the kurtosis, denoted as $\kappa = \langle(E_k(t))^4\rangle/\langle(E_k(t))^2\rangle^2$, where the angular brackets represent the average over different times. The kurtosis $\kappa$ is presented as black squares in Fig.~\ref{fig:scaling}(c), with black dashed lines marking the critical value of $\tau_p^c$, at which $\kappa$ begins to exceed $3$, the kurtosis value corresponding to Gaussian fluctuations.

At large persistent times, $\tau_p>\tau_p^c$, a distinct behavior is observed where $\Gamma_{\rm{T1}}^{\rm irr}$ decreases with increasing $\tau_p$, indicating different relaxation modes when dynamics become intermittent in persistent cells. Within these cells, mechanical equilibrium is readily achieved with ${\bf f}_i^m + {\bf f}_i^a =-{\bf{\nabla}}_ie + v_0\hat{n}_i=0$~\cite{Mandal_nc2020,Berthier_prl2022}, leading to prolonged trapping of the system in the energy minimum of an effective potential $e^{\rm{eff}}=e - \Sigma_iv_0\hat{n}_i\cdot\vec{r}_i$. The break of this mechanical equilibrium relies on the change of $\hat{n}_i$ with a typical time scale of $\tau_p$ as the system slowly evolves its positions. An increase in $\tau_p$ consequently results in a decrease of rearranging events and, as a consequence, a decrease in the irreversible T1 transition rate $\Gamma_{\rm{T1}}^{\rm irr}$. As T1 transitions become rare, the mean velocity magnitude $v=\langle |\mathbf{v}_i|\rangle$ also decreases with increasing $\tau_p$ at $\tau_p>\tau_p^c$ (Fig. S3~\cite{sm}). This contrasts with the observation that $v$ reaches a plateau value close to $v_0=1.0$ at small values of $\tau_p$ (Fig. S3~\cite{sm}). The decrease in both the irreversible T1 transition rate and the mean velocity magnitude contributes to the slow relaxation of the system. This suggests that additional relaxation modes may compensate for this decrease, considering that the final CR $\alpha$ relaxation time is comparable to that at low persistent times.

At high $\tau_p$, the spatial velocity correlation leads to the collective motion of cells with a typical length scale $\xi$. The mechanical equilibrium condition further leads cells to move with a constant velocity, resulting in a long ballistic and super-diffusive region before caging. This introduces a mode where cells relax collectively. Consequently, in Fig. S2~\cite{sm}, we observe that as $\tau_p$ increases, the super-diffusive region extends to a larger time scale, and the subsequent caging plateau starts at later times. These leads to the gradually shrinking of the caging regime gradually shortens with increasing $\tau_p$ for $\tau_p>\tau_p^c$, while this phenomenon is absent at small $\tau_p$ (Fig. S2~\cite{sm}).

Taking into account collective relaxation, we plot the CR $\alpha$ relaxation time in Fig.~\ref{fig:scaling}(d) as a function of $\exp(\xi/\xi_0)\Gamma_{\rm{T1}}^{\rm irr}$, where $\xi_0$ is a constant ranging from approximately $0.7$ to $2.1$ as $\tau_p$ varies from $1.0$ to $10^2$. The data collapse well to a universal function $\tau_{\rm CR}\propto (\exp(\xi/\xi_0)\Gamma_{\rm{T1}}^{\rm irr})^{-\alpha}$ with the dynamical critical exponent $\alpha\simeq1.9$. This collapse spans approximately five orders of magnitude of the CR $\alpha$ relaxation time and a large range of $\tau_p$. At small persistent times, $\tau_p\leq\tau_p^c$, the spatial velocity correlation is absent, and thus $\xi=0$. In these cases, the $\alpha$ relaxation process is solely determined by the irreversible T1 transition rates. In these nearly equilibrated cells, the relaxation process is dominated by the rate of jumps between energy barriers within the energy landscape. These jumps are facilitated by irreversible T1 transitions, and thus, their rates, $\Gamma_{\rm T1}^{\rm irr}$, play a pivotal role in the relaxation dynamics. In contrast, in more persistent cells, irreversible T1 transitions become more infrequent. Cells may shift with a constant velocity collectively with a typical length scale $\xi$ as the mechanical equilibrium condition at short times is attained. This collective persistent motion also contributes to the relaxation dynamics, introducing a dependence of the relaxation time on the additional parameter, $\xi$. These distinct relaxation processes are well described by the universal power-law function, reminiscent of the power-law form of the mode-coupling theory. The divergence of the CR $\alpha$ relaxation time occurs at the mode-coupling transition point, where the irreversible T1 transition rates attain $0$. Persistent motion induces spatial velocity correlation, and triggers collective relaxation, leading to a flatter divergence of $\tau_{\rm CR}$ as irreversible T1 transition nearly vanishes.

In summary, we investigated the liquid-glass transition in confluent monolayers of active epithelial cells using the self-propelled Voronoi model. The glass transition is triggered by decreasing the deformability ($p_0$) and/or the persistent time ($\tau_p$), enabling us to construct a dynamical phase diagram in the $\tau_p$ vs. $p_0$ plane. We demonstrated the influence of system size on dynamics, arising from long-wavelength fluctuations at small persistent times in nearly equilibrated cells and additionally from nonequilibrium effects at large persistent times in active cells. We applied CR measures for dynamical quantities influenced by long-wavelength fluctuations and conducted studies in relatively large systems to ensure negligible finite size effects. While supercooled dynamics occurs at both small and large persistent times, self-propulsions qualitatively alter how the system relaxes and explores its energy landscape. Specifically, we identified spatial velocity correlations and intermittent dynamics associated with intermittent T1 transitions, inducing a distinct relaxation mode in highly persistent cells compared to nearly equilibrated cells. However, these different relaxation modes are described by a universal power-law dependence of CR $\alpha$ relaxation time on the exponential of velocity correlation length times the irreversible T1 transition rate. Our results may inspire novel glass transition theories and stimulate numerical and experimental research into the relaxation dynamics of supercooled active systems.

This work is supported by the National Natural Science Foundation of China (Grants No. 12374204 and No. 12105012).

\bibliographystyle{apsrev4-1}
\input{input.bbl}
\end{document}

%% file: input.bbl
%

%% file: Active_Cell.bbl
\begin{thebibliography}{51}%
\makeatletter
\providecommand \@ifxundefined [1]{%
 \@ifx{#1\undefined}
}%
\providecommand \@ifnum [1]{%
 \ifnum #1\expandafter \@firstoftwo
 \else \expandafter \@secondoftwo
 \fi
}%
\providecommand \@ifx [1]{%
 \ifx #1\expandafter \@firstoftwo
 \else \expandafter \@secondoftwo
 \fi
}%
\providecommand \natexlab [1]{#1}%
\providecommand \enquote  [1]{``#1''}%
\providecommand \bibnamefont  [1]{#1}%
\providecommand \bibfnamefont [1]{#1}%
\providecommand \citenamefont [1]{#1}%
\providecommand \href@noop [0]{\@secondoftwo}%
\providecommand \href [0]{\begingroup \@sanitize@url \@href}%
\providecommand \@href[1]{\@@startlink{#1}\@@href}%
\providecommand \@@href[1]{\endgroup#1\@@endlink}%
\providecommand \@sanitize@url [0]{\catcode `\\12\catcode `\$12\catcode
  `\&12\catcode `\#12\catcode `\^12\catcode `\_12\catcode `\%12\relax}%
\providecommand \@@startlink[1]{}%
\providecommand \@@endlink[0]{}%
\providecommand \url  [0]{\begingroup\@sanitize@url \@url }%
\providecommand \@url [1]{\endgroup\@href {#1}{\urlprefix }}%
\providecommand \urlprefix  [0]{URL }%
\providecommand \Eprint [0]{\href }%
\providecommand \doibase [0]{http://dx.doi.org/}%
\providecommand \selectlanguage [0]{\@gobble}%
\providecommand \bibinfo  [0]{\@secondoftwo}%
\providecommand \bibfield  [0]{\@secondoftwo}%
\providecommand \translation [1]{[#1]}%
\providecommand \BibitemOpen [0]{}%
\providecommand \bibitemStop [0]{}%
\providecommand \bibitemNoStop [0]{.\EOS\space}%
\providecommand \EOS [0]{\spacefactor3000\relax}%
\providecommand \BibitemShut  [1]{\csname bibitem#1\endcsname}%
\let\auto@bib@innerbib\@empty
\bibitem [{\citenamefont {Marchetti}\ \emph {et~al.}(2013)\citenamefont
  {Marchetti}, \citenamefont {Joanny}, \citenamefont {Ramaswamy}, \citenamefont
  {Liverpool}, \citenamefont {Prost}, \citenamefont {Rao},\ and\ \citenamefont
  {Simha}}]{Aditi_Review}%
  \BibitemOpen
  \bibfield  {author} {\bibinfo {author} {\bibfnamefont {M.~C.}\ \bibnamefont
  {Marchetti}}, \bibinfo {author} {\bibfnamefont {J.~F.}\ \bibnamefont
  {Joanny}}, \bibinfo {author} {\bibfnamefont {S.}~\bibnamefont {Ramaswamy}},
  \bibinfo {author} {\bibfnamefont {T.~B.}\ \bibnamefont {Liverpool}}, \bibinfo
  {author} {\bibfnamefont {J.}~\bibnamefont {Prost}}, \bibinfo {author}
  {\bibfnamefont {M.}~\bibnamefont {Rao}}, \ and\ \bibinfo {author}
  {\bibfnamefont {R.~A.}\ \bibnamefont {Simha}},\ }\href@noop {} {\bibfield
  {journal} {\bibinfo  {journal} {Rev. Mod. Phys}\ }\textbf {\bibinfo {volume}
  {85}},\ \bibinfo {pages} {1143} (\bibinfo {year} {2013})}\BibitemShut
  {NoStop}%
\bibitem [{\citenamefont {Bechinger}\ \emph {et~al.}(2016)\citenamefont
  {Bechinger}, \citenamefont {Di~Leonardo}, \citenamefont {Löwen},
  \citenamefont {Reichhardt}, \citenamefont {Volpe},\ and\ \citenamefont
  {Volpe}}]{Bechinger_Review}%
  \BibitemOpen
  \bibfield  {author} {\bibinfo {author} {\bibfnamefont {C.}~\bibnamefont
  {Bechinger}}, \bibinfo {author} {\bibfnamefont {R.}~\bibnamefont
  {Di~Leonardo}}, \bibinfo {author} {\bibfnamefont {H.}~\bibnamefont {Löwen}},
  \bibinfo {author} {\bibfnamefont {C.}~\bibnamefont {Reichhardt}}, \bibinfo
  {author} {\bibfnamefont {G.}~\bibnamefont {Volpe}}, \ and\ \bibinfo {author}
  {\bibfnamefont {G.}~\bibnamefont {Volpe}},\ }\href@noop {} {\bibfield
  {journal} {\bibinfo  {journal} {Rev. Mod. Phys}\ }\textbf {\bibinfo {volume}
  {88}},\ \bibinfo {pages} {045006} (\bibinfo {year} {2016})}\BibitemShut
  {NoStop}%
\bibitem [{\citenamefont {Bi}\ \emph {et~al.}(2016)\citenamefont {Bi},
  \citenamefont {Yang}, \citenamefont {Marchetti},\ and\ \citenamefont
  {Manning}}]{Bi_prx}%
  \BibitemOpen
  \bibfield  {author} {\bibinfo {author} {\bibfnamefont {D.}~\bibnamefont
  {Bi}}, \bibinfo {author} {\bibfnamefont {X.}~\bibnamefont {Yang}}, \bibinfo
  {author} {\bibfnamefont {M.~C.}\ \bibnamefont {Marchetti}}, \ and\ \bibinfo
  {author} {\bibfnamefont {M.~L.}\ \bibnamefont {Manning}},\ }\href@noop {}
  {\bibfield  {journal} {\bibinfo  {journal} {Phys. Rev. X}\ }\textbf {\bibinfo
  {volume} {6}},\ \bibinfo {pages} {021011} (\bibinfo {year}
  {2016})}\BibitemShut {NoStop}%
\bibitem [{\citenamefont {Pasupalak}\ \emph {et~al.}(2020)\citenamefont
  {Pasupalak}, \citenamefont {Yan-Wei}, \citenamefont {Ni},\ and\ \citenamefont
  {Pica~Ciamarra}}]{Anshuman_SM}%
  \BibitemOpen
  \bibfield  {author} {\bibinfo {author} {\bibfnamefont {A.}~\bibnamefont
  {Pasupalak}}, \bibinfo {author} {\bibfnamefont {L.}~\bibnamefont {Yan-Wei}},
  \bibinfo {author} {\bibfnamefont {R.}~\bibnamefont {Ni}}, \ and\ \bibinfo
  {author} {\bibfnamefont {M.}~\bibnamefont {Pica~Ciamarra}},\ }\href@noop {}
  {\bibfield  {journal} {\bibinfo  {journal} {Soft Matter}\ }\textbf {\bibinfo
  {volume} {16}},\ \bibinfo {pages} {3914} (\bibinfo {year}
  {2020})}\BibitemShut {NoStop}%
\bibitem [{\citenamefont {Loewe}\ \emph {et~al.}(2020)\citenamefont {Loewe},
  \citenamefont {Chiang}, \citenamefont {Marenduzzo},\ and\ \citenamefont
  {Marchetti}}]{melting_activecells}%
  \BibitemOpen
  \bibfield  {author} {\bibinfo {author} {\bibfnamefont {B.}~\bibnamefont
  {Loewe}}, \bibinfo {author} {\bibfnamefont {M.}~\bibnamefont {Chiang}},
  \bibinfo {author} {\bibfnamefont {D.}~\bibnamefont {Marenduzzo}}, \ and\
  \bibinfo {author} {\bibfnamefont {M.~C.}\ \bibnamefont {Marchetti}},\
  }\href@noop {} {\bibfield  {journal} {\bibinfo  {journal} {Phys. Rev. Lett.}\
  }\textbf {\bibinfo {volume} {125}},\ \bibinfo {pages} {038003} (\bibinfo
  {year} {2020})}\BibitemShut {NoStop}%
\bibitem [{\citenamefont {Das}\ \emph {et~al.}(2021)\citenamefont {Das},
  \citenamefont {Sastry},\ and\ \citenamefont {Bi}}]{Bi_prx2021}%
  \BibitemOpen
  \bibfield  {author} {\bibinfo {author} {\bibfnamefont {A.}~\bibnamefont
  {Das}}, \bibinfo {author} {\bibfnamefont {S.}~\bibnamefont {Sastry}}, \ and\
  \bibinfo {author} {\bibfnamefont {D.}~\bibnamefont {Bi}},\ }\href@noop {}
  {\bibfield  {journal} {\bibinfo  {journal} {Phys. Rev. X}\ }\textbf {\bibinfo
  {volume} {11}},\ \bibinfo {pages} {041037} (\bibinfo {year}
  {2021})}\BibitemShut {NoStop}%
\bibitem [{\citenamefont {Kim}\ \emph {et~al.}(2013)\citenamefont {Kim},
  \citenamefont {Serra-Picamal}, \citenamefont {Tambe}, \citenamefont {Zhou},
  \citenamefont {Park}, \citenamefont {Sadati}, \citenamefont {Park},
  \citenamefont {Krishnan}, \citenamefont {Gweon}, \citenamefont {Millet},
  \citenamefont {Butler}, \citenamefont {Trepat},\ and\ \citenamefont
  {Fredberg}}]{Kim_NM}%
  \BibitemOpen
  \bibfield  {author} {\bibinfo {author} {\bibfnamefont {J.~H.}\ \bibnamefont
  {Kim}}, \bibinfo {author} {\bibfnamefont {X.}~\bibnamefont {Serra-Picamal}},
  \bibinfo {author} {\bibfnamefont {D.~T.}\ \bibnamefont {Tambe}}, \bibinfo
  {author} {\bibfnamefont {E.~H.}\ \bibnamefont {Zhou}}, \bibinfo {author}
  {\bibfnamefont {C.~Y.}\ \bibnamefont {Park}}, \bibinfo {author}
  {\bibfnamefont {M.}~\bibnamefont {Sadati}}, \bibinfo {author} {\bibfnamefont
  {J.-A.}\ \bibnamefont {Park}}, \bibinfo {author} {\bibfnamefont
  {R.}~\bibnamefont {Krishnan}}, \bibinfo {author} {\bibfnamefont
  {B.}~\bibnamefont {Gweon}}, \bibinfo {author} {\bibfnamefont
  {E.}~\bibnamefont {Millet}}, \bibinfo {author} {\bibfnamefont {J.~P.}\
  \bibnamefont {Butler}}, \bibinfo {author} {\bibfnamefont {X.}~\bibnamefont
  {Trepat}}, \ and\ \bibinfo {author} {\bibfnamefont {J.~J.}\ \bibnamefont
  {Fredberg}},\ }\href@noop {} {\bibfield  {journal} {\bibinfo  {journal} {Nat.
  Mater.}\ }\textbf {\bibinfo {volume} {12}},\ \bibinfo {pages} {856} (\bibinfo
  {year} {2013})}\BibitemShut {NoStop}%
\bibitem [{\citenamefont {Garcia}\ \emph {et~al.}(2015)\citenamefont {Garcia},
  \citenamefont {Hannezo}, \citenamefont {Elgeti}, \citenamefont {Joanny},
  \citenamefont {Silberzan},\ and\ \citenamefont {Gov}}]{Simon_pnas}%
  \BibitemOpen
  \bibfield  {author} {\bibinfo {author} {\bibfnamefont {S.}~\bibnamefont
  {Garcia}}, \bibinfo {author} {\bibfnamefont {E.}~\bibnamefont {Hannezo}},
  \bibinfo {author} {\bibfnamefont {J.}~\bibnamefont {Elgeti}}, \bibinfo
  {author} {\bibfnamefont {J.-F.}\ \bibnamefont {Joanny}}, \bibinfo {author}
  {\bibfnamefont {P.}~\bibnamefont {Silberzan}}, \ and\ \bibinfo {author}
  {\bibfnamefont {N.~S.}\ \bibnamefont {Gov}},\ }\href@noop {} {\bibfield
  {journal} {\bibinfo  {journal} {Proc. Natl. Acad. Sci. U.S.A.}\ }\textbf
  {\bibinfo {volume} {112}},\ \bibinfo {pages} {15314} (\bibinfo {year}
  {2015})}\BibitemShut {NoStop}%
\bibitem [{\citenamefont {Armengol-Collado}\ \emph {et~al.}(2023)\citenamefont
  {Armengol-Collado}, \citenamefont {Carenza}, \citenamefont {Eckert},
  \citenamefont {Krommydas},\ and\ \citenamefont {Giomi}}]{Luca_NP}%
  \BibitemOpen
  \bibfield  {author} {\bibinfo {author} {\bibfnamefont {J.-M.}\ \bibnamefont
  {Armengol-Collado}}, \bibinfo {author} {\bibfnamefont {L.~N.}\ \bibnamefont
  {Carenza}}, \bibinfo {author} {\bibfnamefont {J.}~\bibnamefont {Eckert}},
  \bibinfo {author} {\bibfnamefont {D.}~\bibnamefont {Krommydas}}, \ and\
  \bibinfo {author} {\bibfnamefont {L.}~\bibnamefont {Giomi}},\ }\href@noop {}
  {\bibfield  {journal} {\bibinfo  {journal} {Nat. Phys.}\ } (\bibinfo {year}
  {2023})}\BibitemShut {NoStop}%
\bibitem [{\citenamefont {Barton}\ \emph {et~al.}(2017)\citenamefont {Barton},
  \citenamefont {Henkes}, \citenamefont {Weijer},\ and\ \citenamefont
  {Sknepnek}}]{Barton2017}%
  \BibitemOpen
  \bibfield  {author} {\bibinfo {author} {\bibfnamefont {D.~L.}\ \bibnamefont
  {Barton}}, \bibinfo {author} {\bibfnamefont {S.}~\bibnamefont {Henkes}},
  \bibinfo {author} {\bibfnamefont {C.~J.}\ \bibnamefont {Weijer}}, \ and\
  \bibinfo {author} {\bibfnamefont {R.}~\bibnamefont {Sknepnek}},\ }\href@noop
  {} {\bibfield  {journal} {\bibinfo  {journal} {PLoS Comput. Biol.}\ }\textbf
  {\bibinfo {volume} {13}},\ \bibinfo {pages} {34} (\bibinfo {year}
  {2017})}\BibitemShut {NoStop}%
\bibitem [{\citenamefont {Basan}\ \emph {et~al.}(2013)\citenamefont {Basan},
  \citenamefont {Elgeti}, \citenamefont {Hannezo}, \citenamefont {Rappel},\
  and\ \citenamefont {Levine}}]{Levine_PNAS}%
  \BibitemOpen
  \bibfield  {author} {\bibinfo {author} {\bibfnamefont {M.}~\bibnamefont
  {Basan}}, \bibinfo {author} {\bibfnamefont {J.}~\bibnamefont {Elgeti}},
  \bibinfo {author} {\bibfnamefont {E.}~\bibnamefont {Hannezo}}, \bibinfo
  {author} {\bibfnamefont {W.-J.}\ \bibnamefont {Rappel}}, \ and\ \bibinfo
  {author} {\bibfnamefont {H.}~\bibnamefont {Levine}},\ }\href@noop {}
  {\bibfield  {journal} {\bibinfo  {journal} {Proc. Natl. Acad. Sci. U.S.A.}\
  }\textbf {\bibinfo {volume} {110}},\ \bibinfo {pages} {2452} (\bibinfo {year}
  {2013})}\BibitemShut {NoStop}%
\bibitem [{\citenamefont {Poujade}\ \emph {et~al.}(2007)\citenamefont
  {Poujade}, \citenamefont {Grasland-Mongrain}, \citenamefont {Hertzog},
  \citenamefont {Jouanneau}, \citenamefont {Chavrier}, \citenamefont {Ladoux},
  \citenamefont {Buguin},\ and\ \citenamefont {Silberzan}}]{Silberzan_pnas}%
  \BibitemOpen
  \bibfield  {author} {\bibinfo {author} {\bibfnamefont {M.}~\bibnamefont
  {Poujade}}, \bibinfo {author} {\bibfnamefont {E.}~\bibnamefont
  {Grasland-Mongrain}}, \bibinfo {author} {\bibfnamefont {A.}~\bibnamefont
  {Hertzog}}, \bibinfo {author} {\bibfnamefont {J.}~\bibnamefont {Jouanneau}},
  \bibinfo {author} {\bibfnamefont {P.}~\bibnamefont {Chavrier}}, \bibinfo
  {author} {\bibfnamefont {B.}~\bibnamefont {Ladoux}}, \bibinfo {author}
  {\bibfnamefont {A.}~\bibnamefont {Buguin}}, \ and\ \bibinfo {author}
  {\bibfnamefont {P.}~\bibnamefont {Silberzan}},\ }\href@noop {} {\bibfield
  {journal} {\bibinfo  {journal} {Proc. Natl. Acad. Sci. U.S.A.}\ }\textbf
  {\bibinfo {volume} {104}},\ \bibinfo {pages} {15988} (\bibinfo {year}
  {2007})}\BibitemShut {NoStop}%
\bibitem [{\citenamefont {Manli}\ \emph {et~al.}(2012)\citenamefont {Manli},
  \citenamefont {David},\ and\ \citenamefont {Cornelis}}]{Cornelis}%
  \BibitemOpen
  \bibfield  {author} {\bibinfo {author} {\bibfnamefont {C.}~\bibnamefont
  {Manli}}, \bibinfo {author} {\bibfnamefont {H.}~\bibnamefont {David}}, \ and\
  \bibinfo {author} {\bibfnamefont {J.~W.}\ \bibnamefont {Cornelis}},\
  }\href@noop {} {\bibfield  {journal} {\bibinfo  {journal} {Curr. Genomics}\
  }\textbf {\bibinfo {volume} {13}},\ \bibinfo {pages} {267} (\bibinfo {year}
  {2012})}\BibitemShut {NoStop}%
\bibitem [{\citenamefont {Schötz}\ \emph {et~al.}(2013)\citenamefont
  {Schötz}, \citenamefont {Lanio}, \citenamefont {Talbot},\ and\ \citenamefont
  {Manning}}]{Manning2013}%
  \BibitemOpen
  \bibfield  {author} {\bibinfo {author} {\bibfnamefont {E.-M.}\ \bibnamefont
  {Schötz}}, \bibinfo {author} {\bibfnamefont {M.}~\bibnamefont {Lanio}},
  \bibinfo {author} {\bibfnamefont {J.~A.}\ \bibnamefont {Talbot}}, \ and\
  \bibinfo {author} {\bibfnamefont {M.~L.}\ \bibnamefont {Manning}},\
  }\href@noop {} {\bibfield  {journal} {\bibinfo  {journal} {J. R. Soc.
  Interface}\ }\textbf {\bibinfo {volume} {10}} (\bibinfo {year}
  {2013})}\BibitemShut {NoStop}%
\bibitem [{\citenamefont {Angelini}\ \emph {et~al.}(2011)\citenamefont
  {Angelini}, \citenamefont {Hannezo}, \citenamefont {Trepat}, \citenamefont
  {Marquez}, \citenamefont {Fredberg},\ and\ \citenamefont
  {Weitz}}]{Weitz_PNAS2011}%
  \BibitemOpen
  \bibfield  {author} {\bibinfo {author} {\bibfnamefont {T.~E.}\ \bibnamefont
  {Angelini}}, \bibinfo {author} {\bibfnamefont {E.}~\bibnamefont {Hannezo}},
  \bibinfo {author} {\bibfnamefont {X.}~\bibnamefont {Trepat}}, \bibinfo
  {author} {\bibfnamefont {M.}~\bibnamefont {Marquez}}, \bibinfo {author}
  {\bibfnamefont {J.~J.}\ \bibnamefont {Fredberg}}, \ and\ \bibinfo {author}
  {\bibfnamefont {D.~A.}\ \bibnamefont {Weitz}},\ }\href@noop {} {\bibfield
  {journal} {\bibinfo  {journal} {Proc. Natl. Acad. Sci. U.S.A.}\ }\textbf
  {\bibinfo {volume} {108}},\ \bibinfo {pages} {4714} (\bibinfo {year}
  {2011})}\BibitemShut {NoStop}%
\bibitem [{\citenamefont {Jordan}\ \emph {et~al.}(2011)\citenamefont {Jordan},
  \citenamefont {Johnson},\ and\ \citenamefont {Abell}}]{EMT1}%
  \BibitemOpen
  \bibfield  {author} {\bibinfo {author} {\bibfnamefont {N.~V.}\ \bibnamefont
  {Jordan}}, \bibinfo {author} {\bibfnamefont {G.~L.}\ \bibnamefont {Johnson}},
  \ and\ \bibinfo {author} {\bibfnamefont {A.~N.}\ \bibnamefont {Abell}},\
  }\href@noop {} {\bibfield  {journal} {\bibinfo  {journal} {Cell Cycle}\
  }\textbf {\bibinfo {volume} {10}},\ \bibinfo {pages} {2865} (\bibinfo {year}
  {2011})}\BibitemShut {NoStop}%
\bibitem [{\citenamefont {Tanaka}\ and\ \citenamefont {Ogishima}(2015)}]{EMT2}%
  \BibitemOpen
  \bibfield  {author} {\bibinfo {author} {\bibfnamefont {H.}~\bibnamefont
  {Tanaka}}\ and\ \bibinfo {author} {\bibfnamefont {S.}~\bibnamefont
  {Ogishima}},\ }\href@noop {} {\bibfield  {journal} {\bibinfo  {journal} {J.
  Mol. Cell Biol.}\ }\textbf {\bibinfo {volume} {7}},\ \bibinfo {pages} {253}
  (\bibinfo {year} {2015})}\BibitemShut {NoStop}%
\bibitem [{\citenamefont {Kalluri}\ and\ \citenamefont
  {Weinberg}(2009)}]{EMT3}%
  \BibitemOpen
  \bibfield  {author} {\bibinfo {author} {\bibfnamefont {R.}~\bibnamefont
  {Kalluri}}\ and\ \bibinfo {author} {\bibfnamefont {R.~A.}\ \bibnamefont
  {Weinberg}},\ }\href@noop {} {\bibfield  {journal} {\bibinfo  {journal} {J.
  Clin. Invest.}\ }\textbf {\bibinfo {volume} {119}},\ \bibinfo {pages} {1420}
  (\bibinfo {year} {2009})}\BibitemShut {NoStop}%
\bibitem [{\citenamefont {Park}\ \emph {et~al.}(2015)\citenamefont {Park},
  \citenamefont {Kim}, \citenamefont {Bi}, \citenamefont {Mitchel},
  \citenamefont {Qazvini}, \citenamefont {Tantisira}, \citenamefont {Park},
  \citenamefont {McGill}, \citenamefont {Kim}, \citenamefont {Gweon},
  \citenamefont {Notbohm}, \citenamefont {Steward~Jr}, \citenamefont {Burger},
  \citenamefont {Randell}, \citenamefont {Kho}, \citenamefont {Tambe},
  \citenamefont {Hardin}, \citenamefont {Shore}, \citenamefont {Israel},
  \citenamefont {Weitz}, \citenamefont {Tschumperlin}, \citenamefont {Henske},
  \citenamefont {Weiss}, \citenamefont {Manning}, \citenamefont {Butler},
  \citenamefont {Drazen},\ and\ \citenamefont {Fredberg}}]{Park_NM}%
  \BibitemOpen
  \bibfield  {author} {\bibinfo {author} {\bibfnamefont {J.-A.}\ \bibnamefont
  {Park}}, \bibinfo {author} {\bibfnamefont {J.~H.}\ \bibnamefont {Kim}},
  \bibinfo {author} {\bibfnamefont {D.}~\bibnamefont {Bi}}, \bibinfo {author}
  {\bibfnamefont {J.~A.}\ \bibnamefont {Mitchel}}, \bibinfo {author}
  {\bibfnamefont {N.~T.}\ \bibnamefont {Qazvini}}, \bibinfo {author}
  {\bibfnamefont {K.}~\bibnamefont {Tantisira}}, \bibinfo {author}
  {\bibfnamefont {C.~Y.}\ \bibnamefont {Park}}, \bibinfo {author}
  {\bibfnamefont {M.}~\bibnamefont {McGill}}, \bibinfo {author} {\bibfnamefont
  {S.-H.}\ \bibnamefont {Kim}}, \bibinfo {author} {\bibfnamefont
  {B.}~\bibnamefont {Gweon}}, \bibinfo {author} {\bibfnamefont
  {J.}~\bibnamefont {Notbohm}}, \bibinfo {author} {\bibfnamefont
  {R.}~\bibnamefont {Steward~Jr}}, \bibinfo {author} {\bibfnamefont
  {S.}~\bibnamefont {Burger}}, \bibinfo {author} {\bibfnamefont {S.~H.}\
  \bibnamefont {Randell}}, \bibinfo {author} {\bibfnamefont {A.~T.}\
  \bibnamefont {Kho}}, \bibinfo {author} {\bibfnamefont {D.~T.}\ \bibnamefont
  {Tambe}}, \bibinfo {author} {\bibfnamefont {C.}~\bibnamefont {Hardin}},
  \bibinfo {author} {\bibfnamefont {S.~A.}\ \bibnamefont {Shore}}, \bibinfo
  {author} {\bibfnamefont {E.}~\bibnamefont {Israel}}, \bibinfo {author}
  {\bibfnamefont {D.~A.}\ \bibnamefont {Weitz}}, \bibinfo {author}
  {\bibfnamefont {D.~J.}\ \bibnamefont {Tschumperlin}}, \bibinfo {author}
  {\bibfnamefont {E.~P.}\ \bibnamefont {Henske}}, \bibinfo {author}
  {\bibfnamefont {S.~T.}\ \bibnamefont {Weiss}}, \bibinfo {author}
  {\bibfnamefont {M.~L.}\ \bibnamefont {Manning}}, \bibinfo {author}
  {\bibfnamefont {J.~P.}\ \bibnamefont {Butler}}, \bibinfo {author}
  {\bibfnamefont {J.~M.}\ \bibnamefont {Drazen}}, \ and\ \bibinfo {author}
  {\bibfnamefont {J.~J.}\ \bibnamefont {Fredberg}},\ }\href@noop {} {\bibfield
  {journal} {\bibinfo  {journal} {Nat. Mater.}\ }\textbf {\bibinfo {volume}
  {14}},\ \bibinfo {pages} {1040} (\bibinfo {year} {2015})}\BibitemShut
  {NoStop}%
\bibitem [{\citenamefont {Puliafito}\ \emph {et~al.}(2012)\citenamefont
  {Puliafito}, \citenamefont {Hufnagel}, \citenamefont {Neveu}, \citenamefont
  {Streichan}, \citenamefont {Sigal}, \citenamefont {Fygenson},\ and\
  \citenamefont {Shraiman}}]{Puliafito_PNAS}%
  \BibitemOpen
  \bibfield  {author} {\bibinfo {author} {\bibfnamefont {A.}~\bibnamefont
  {Puliafito}}, \bibinfo {author} {\bibfnamefont {L.}~\bibnamefont {Hufnagel}},
  \bibinfo {author} {\bibfnamefont {P.}~\bibnamefont {Neveu}}, \bibinfo
  {author} {\bibfnamefont {S.}~\bibnamefont {Streichan}}, \bibinfo {author}
  {\bibfnamefont {A.}~\bibnamefont {Sigal}}, \bibinfo {author} {\bibfnamefont
  {D.~K.}\ \bibnamefont {Fygenson}}, \ and\ \bibinfo {author} {\bibfnamefont
  {B.~I.}\ \bibnamefont {Shraiman}},\ }\href@noop {} {\bibfield  {journal}
  {\bibinfo  {journal} {Proc. Natl. Acad. Sci. U.S.A.}\ }\textbf {\bibinfo
  {volume} {109}},\ \bibinfo {pages} {739} (\bibinfo {year}
  {2012})}\BibitemShut {NoStop}%
\bibitem [{\citenamefont {Angelini}\ \emph {et~al.}(2010)\citenamefont
  {Angelini}, \citenamefont {Hannezo}, \citenamefont {Trepat}, \citenamefont
  {Fredberg},\ and\ \citenamefont {Weitz}}]{Angelini_prl}%
  \BibitemOpen
  \bibfield  {author} {\bibinfo {author} {\bibfnamefont {T.~E.}\ \bibnamefont
  {Angelini}}, \bibinfo {author} {\bibfnamefont {E.}~\bibnamefont {Hannezo}},
  \bibinfo {author} {\bibfnamefont {X.}~\bibnamefont {Trepat}}, \bibinfo
  {author} {\bibfnamefont {J.~J.}\ \bibnamefont {Fredberg}}, \ and\ \bibinfo
  {author} {\bibfnamefont {D.~A.}\ \bibnamefont {Weitz}},\ }\href@noop {}
  {\bibfield  {journal} {\bibinfo  {journal} {Phys. Rev. Lett.}\ }\textbf
  {\bibinfo {volume} {104}},\ \bibinfo {pages} {168104} (\bibinfo {year}
  {2010})}\BibitemShut {NoStop}%
\bibitem [{\citenamefont {Sussman}\ \emph {et~al.}(2018)\citenamefont
  {Sussman}, \citenamefont {Paoluzzi}, \citenamefont {Cristina~Marchetti},\
  and\ \citenamefont {Lisa~Manning}}]{Manning_EPL}%
  \BibitemOpen
  \bibfield  {author} {\bibinfo {author} {\bibfnamefont {D.~M.}\ \bibnamefont
  {Sussman}}, \bibinfo {author} {\bibfnamefont {M.}~\bibnamefont {Paoluzzi}},
  \bibinfo {author} {\bibfnamefont {M.}~\bibnamefont {Cristina~Marchetti}}, \
  and\ \bibinfo {author} {\bibfnamefont {M.}~\bibnamefont {Lisa~Manning}},\
  }\href@noop {} {\bibfield  {journal} {\bibinfo  {journal} {Europhys. Lett.}\
  }\textbf {\bibinfo {volume} {121}},\ \bibinfo {pages} {36001} (\bibinfo
  {year} {2018})}\BibitemShut {NoStop}%
\bibitem [{\citenamefont {Li}\ \emph {et~al.}(2021)\citenamefont {Li},
  \citenamefont {Wei}, \citenamefont {Paoluzzi},\ and\ \citenamefont
  {Ciamarra}}]{ywli_pre}%
  \BibitemOpen
  \bibfield  {author} {\bibinfo {author} {\bibfnamefont {Y.-W.}\ \bibnamefont
  {Li}}, \bibinfo {author} {\bibfnamefont {L.~L.~Y.}\ \bibnamefont {Wei}},
  \bibinfo {author} {\bibfnamefont {M.}~\bibnamefont {Paoluzzi}}, \ and\
  \bibinfo {author} {\bibfnamefont {M.~P.}\ \bibnamefont {Ciamarra}},\
  }\href@noop {} {\bibfield  {journal} {\bibinfo  {journal} {Phys. Rev. E}\
  }\textbf {\bibinfo {volume} {103}},\ \bibinfo {pages} {022607} (\bibinfo
  {year} {2021})}\BibitemShut {NoStop}%
\bibitem [{\citenamefont {Mandal}\ \emph {et~al.}(2020)\citenamefont {Mandal},
  \citenamefont {Bhuyan}, \citenamefont {Chaudhuri}, \citenamefont {Dasgupta},\
  and\ \citenamefont {Rao}}]{Mandal_nc2020}%
  \BibitemOpen
  \bibfield  {author} {\bibinfo {author} {\bibfnamefont {R.}~\bibnamefont
  {Mandal}}, \bibinfo {author} {\bibfnamefont {P.~J.}\ \bibnamefont {Bhuyan}},
  \bibinfo {author} {\bibfnamefont {P.}~\bibnamefont {Chaudhuri}}, \bibinfo
  {author} {\bibfnamefont {C.}~\bibnamefont {Dasgupta}}, \ and\ \bibinfo
  {author} {\bibfnamefont {M.}~\bibnamefont {Rao}},\ }\href@noop {} {\bibfield
  {journal} {\bibinfo  {journal} {Nat. Commun.}\ }\textbf {\bibinfo {volume}
  {11}},\ \bibinfo {pages} {2581} (\bibinfo {year} {2020})}\BibitemShut
  {NoStop}%
\bibitem [{\citenamefont {Keta}\ \emph {et~al.}(2022)\citenamefont {Keta},
  \citenamefont {Jack},\ and\ \citenamefont {Berthier}}]{Berthier_prl2022}%
  \BibitemOpen
  \bibfield  {author} {\bibinfo {author} {\bibfnamefont {Y.-E.}\ \bibnamefont
  {Keta}}, \bibinfo {author} {\bibfnamefont {R.~L.}\ \bibnamefont {Jack}}, \
  and\ \bibinfo {author} {\bibfnamefont {L.}~\bibnamefont {Berthier}},\
  }\href@noop {} {\bibfield  {journal} {\bibinfo  {journal} {Phys. Rev. Lett.}\
  }\textbf {\bibinfo {volume} {129}},\ \bibinfo {pages} {048002} (\bibinfo
  {year} {2022})}\BibitemShut {NoStop}%
\bibitem [{\citenamefont {Keta}\ \emph {et~al.}(2023)\citenamefont {Keta},
  \citenamefont {Mandal}, \citenamefont {Sollich}, \citenamefont {Jack},\ and\
  \citenamefont {Berthier}}]{Berthier_sm2023}%
  \BibitemOpen
  \bibfield  {author} {\bibinfo {author} {\bibfnamefont {Y.-E.}\ \bibnamefont
  {Keta}}, \bibinfo {author} {\bibfnamefont {R.}~\bibnamefont {Mandal}},
  \bibinfo {author} {\bibfnamefont {P.}~\bibnamefont {Sollich}}, \bibinfo
  {author} {\bibfnamefont {R.~L.}\ \bibnamefont {Jack}}, \ and\ \bibinfo
  {author} {\bibfnamefont {L.}~\bibnamefont {Berthier}},\ }\href@noop {}
  {\bibfield  {journal} {\bibinfo  {journal} {Soft Matter}\ }\textbf {\bibinfo
  {volume} {19}},\ \bibinfo {pages} {3871} (\bibinfo {year}
  {2023})}\BibitemShut {NoStop}%
\bibitem [{\citenamefont {Brodu}\ \emph {et~al.}(2015)\citenamefont {Brodu},
  \citenamefont {Dijksman},\ and\ \citenamefont {Behringer}}]{dense_granular}%
  \BibitemOpen
  \bibfield  {author} {\bibinfo {author} {\bibfnamefont {N.}~\bibnamefont
  {Brodu}}, \bibinfo {author} {\bibfnamefont {J.~A.}\ \bibnamefont {Dijksman}},
  \ and\ \bibinfo {author} {\bibfnamefont {R.~P.}\ \bibnamefont {Behringer}},\
  }\href@noop {} {\bibfield  {journal} {\bibinfo  {journal} {Phys. Rev. E}\
  }\textbf {\bibinfo {volume} {91}},\ \bibinfo {pages} {032201} (\bibinfo
  {year} {2015})}\BibitemShut {NoStop}%
\bibitem [{\citenamefont {Hohler}\ and\ \citenamefont
  {Cohen-Addad}(2017)}]{soft_jammed}%
  \BibitemOpen
  \bibfield  {author} {\bibinfo {author} {\bibfnamefont {R.}~\bibnamefont
  {Hohler}}\ and\ \bibinfo {author} {\bibfnamefont {S.}~\bibnamefont
  {Cohen-Addad}},\ }\href@noop {} {\bibfield  {journal} {\bibinfo  {journal}
  {Soft Matter}\ }\textbf {\bibinfo {volume} {13}},\ \bibinfo {pages} {1371}
  (\bibinfo {year} {2017})}\BibitemShut {NoStop}%
\bibitem [{\citenamefont {Christos}\ \emph {et~al.}(2003)\citenamefont
  {Christos}, \citenamefont {Norman}, \citenamefont {Arben},\ and\
  \citenamefont {Hartmut}}]{polyelectrolyte_star}%
  \BibitemOpen
  \bibfield  {author} {\bibinfo {author} {\bibfnamefont {N.~L.}\ \bibnamefont
  {Christos}}, \bibinfo {author} {\bibfnamefont {H.}~\bibnamefont {Norman}},
  \bibinfo {author} {\bibfnamefont {J.}~\bibnamefont {Arben}}, \ and\ \bibinfo
  {author} {\bibfnamefont {L.}~\bibnamefont {Hartmut}},\ }\href@noop {}
  {\bibfield  {journal} {\bibinfo  {journal} {J. Phys.: Condens. Matter}\
  }\textbf {\bibinfo {volume} {15}},\ \bibinfo {pages} {S233} (\bibinfo {year}
  {2003})}\BibitemShut {NoStop}%
\bibitem [{\citenamefont {Denton}(2003)}]{microgels}%
  \BibitemOpen
  \bibfield  {author} {\bibinfo {author} {\bibfnamefont {A.~R.}\ \bibnamefont
  {Denton}},\ }\href@noop {} {\bibfield  {journal} {\bibinfo  {journal} {Phys.
  Rev. E}\ }\textbf {\bibinfo {volume} {67}},\ \bibinfo {pages} {011804}
  (\bibinfo {year} {2003})}\BibitemShut {NoStop}%
\bibitem [{\citenamefont {Watzlawek}\ \emph {et~al.}(1999)\citenamefont
  {Watzlawek}, \citenamefont {Likos},\ and\ \citenamefont
  {L\"{o}wen}}]{Star_polymer}%
  \BibitemOpen
  \bibfield  {author} {\bibinfo {author} {\bibfnamefont {M.}~\bibnamefont
  {Watzlawek}}, \bibinfo {author} {\bibfnamefont {C.~N.}\ \bibnamefont
  {Likos}}, \ and\ \bibinfo {author} {\bibfnamefont {H.}~\bibnamefont
  {L\"{o}wen}},\ }\href@noop {} {\bibfield  {journal} {\bibinfo  {journal}
  {Phys. Rev. Lett.}\ }\textbf {\bibinfo {volume} {82}},\ \bibinfo {pages}
  {5289} (\bibinfo {year} {1999})}\BibitemShut {NoStop}%
\bibitem [{\citenamefont {von Ferber}\ \emph {et~al.}(2000)\citenamefont {von
  Ferber}, \citenamefont {Jusufi}, \citenamefont {Likos}, \citenamefont
  {L\"{o}wen},\ and\ \citenamefont {Watzlawek}}]{Likos_star}%
  \BibitemOpen
  \bibfield  {author} {\bibinfo {author} {\bibfnamefont {C.}~\bibnamefont {von
  Ferber}}, \bibinfo {author} {\bibfnamefont {A.}~\bibnamefont {Jusufi}},
  \bibinfo {author} {\bibfnamefont {C.~N.}\ \bibnamefont {Likos}}, \bibinfo
  {author} {\bibfnamefont {H.}~\bibnamefont {L\"{o}wen}}, \ and\ \bibinfo
  {author} {\bibfnamefont {M.}~\bibnamefont {Watzlawek}},\ }\href@noop {}
  {\bibfield  {journal} {\bibinfo  {journal} {Eur. Phys. J. E}\ }\textbf
  {\bibinfo {volume} {2}},\ \bibinfo {pages} {311} (\bibinfo {year}
  {2000})}\BibitemShut {NoStop}%
\bibitem [{\citenamefont {Terao}(2006)}]{Dendrimer}%
  \BibitemOpen
  \bibfield  {author} {\bibinfo {author} {\bibfnamefont {T.}~\bibnamefont
  {Terao}},\ }\href@noop {} {\bibfield  {journal} {\bibinfo  {journal} {Mol.
  Phys.}\ }\textbf {\bibinfo {volume} {104}},\ \bibinfo {pages} {2507}
  (\bibinfo {year} {2006})}\BibitemShut {NoStop}%
\bibitem [{\citenamefont {Weaire}\ and\ \citenamefont
  {Rivier}(1984)}]{Rivier_1984}%
  \BibitemOpen
  \bibfield  {author} {\bibinfo {author} {\bibfnamefont {D.}~\bibnamefont
  {Weaire}}\ and\ \bibinfo {author} {\bibfnamefont {N.}~\bibnamefont
  {Rivier}},\ }\href@noop {} {\bibfield  {journal} {\bibinfo  {journal}
  {Contemp. Phys.}\ }\textbf {\bibinfo {volume} {25}},\ \bibinfo {pages} {59}
  (\bibinfo {year} {1984})}\BibitemShut {NoStop}%
\bibitem [{\citenamefont {Staple}\ \emph {et~al.}(2010)\citenamefont {Staple},
  \citenamefont {Farhadifar}, \citenamefont {R\"{o}per}, \citenamefont
  {Aigouy}, \citenamefont {Eaton},\ and\ \citenamefont
  {J\"{u}licher}}]{Staple2010}%
  \BibitemOpen
  \bibfield  {author} {\bibinfo {author} {\bibfnamefont {D.~B.}\ \bibnamefont
  {Staple}}, \bibinfo {author} {\bibfnamefont {R.}~\bibnamefont {Farhadifar}},
  \bibinfo {author} {\bibfnamefont {J.-C.}\ \bibnamefont {R\"{o}per}}, \bibinfo
  {author} {\bibfnamefont {B.}~\bibnamefont {Aigouy}}, \bibinfo {author}
  {\bibfnamefont {S.}~\bibnamefont {Eaton}}, \ and\ \bibinfo {author}
  {\bibfnamefont {F.}~\bibnamefont {J\"{u}licher}},\ }\href@noop {} {\bibfield
  {journal} {\bibinfo  {journal} {Eur. Phys. J. E}\ }\textbf {\bibinfo {volume}
  {33}},\ \bibinfo {pages} {117} (\bibinfo {year} {2010})}\BibitemShut
  {NoStop}%
\bibitem [{\citenamefont {Bi}\ \emph {et~al.}(2014)\citenamefont {Bi},
  \citenamefont {Lopez}, \citenamefont {Schwarz},\ and\ \citenamefont
  {Manning}}]{Bi_softMatter}%
  \BibitemOpen
  \bibfield  {author} {\bibinfo {author} {\bibfnamefont {D.}~\bibnamefont
  {Bi}}, \bibinfo {author} {\bibfnamefont {J.~H.}\ \bibnamefont {Lopez}},
  \bibinfo {author} {\bibfnamefont {J.~M.}\ \bibnamefont {Schwarz}}, \ and\
  \bibinfo {author} {\bibfnamefont {M.~L.}\ \bibnamefont {Manning}},\
  }\href@noop {} {\bibfield  {journal} {\bibinfo  {journal} {Soft Matter}\
  }\textbf {\bibinfo {volume} {10}},\ \bibinfo {pages} {1885} (\bibinfo {year}
  {2014})}\BibitemShut {NoStop}%
\bibitem [{\citenamefont {Farhadifar}\ \emph {et~al.}(2007)\citenamefont
  {Farhadifar}, \citenamefont {R\"{o}per}, \citenamefont {Aigouy},
  \citenamefont {Eaton},\ and\ \citenamefont {J\"{u}licher}}]{Frank_vertex}%
  \BibitemOpen
  \bibfield  {author} {\bibinfo {author} {\bibfnamefont {R.}~\bibnamefont
  {Farhadifar}}, \bibinfo {author} {\bibfnamefont {J.-C.}\ \bibnamefont
  {R\"{o}per}}, \bibinfo {author} {\bibfnamefont {B.}~\bibnamefont {Aigouy}},
  \bibinfo {author} {\bibfnamefont {S.}~\bibnamefont {Eaton}}, \ and\ \bibinfo
  {author} {\bibfnamefont {F.}~\bibnamefont {J\"{u}licher}},\ }\href@noop {}
  {\bibfield  {journal} {\bibinfo  {journal} {Curr. Biol.}\ }\textbf {\bibinfo
  {volume} {17}},\ \bibinfo {pages} {2095} (\bibinfo {year}
  {2007})}\BibitemShut {NoStop}%
\bibitem [{\citenamefont {Bi}\ \emph {et~al.}(2015)\citenamefont {Bi},
  \citenamefont {Lopez}, \citenamefont {Schwarz},\ and\ \citenamefont
  {Manning}}]{Bi_np2015}%
  \BibitemOpen
  \bibfield  {author} {\bibinfo {author} {\bibfnamefont {D.}~\bibnamefont
  {Bi}}, \bibinfo {author} {\bibfnamefont {J.~H.}\ \bibnamefont {Lopez}},
  \bibinfo {author} {\bibfnamefont {J.~M.}\ \bibnamefont {Schwarz}}, \ and\
  \bibinfo {author} {\bibfnamefont {M.~L.}\ \bibnamefont {Manning}},\
  }\href@noop {} {\bibfield  {journal} {\bibinfo  {journal} {Nat. Phys.}\
  }\textbf {\bibinfo {volume} {11}},\ \bibinfo {pages} {1074} (\bibinfo {year}
  {2015})}\BibitemShut {NoStop}%
\bibitem [{\citenamefont {Manning}\ \emph {et~al.}(2010)\citenamefont
  {Manning}, \citenamefont {Foty}, \citenamefont {Steinberg},\ and\
  \citenamefont {Schoetz}}]{Manning1}%
  \BibitemOpen
  \bibfield  {author} {\bibinfo {author} {\bibfnamefont {M.~L.}\ \bibnamefont
  {Manning}}, \bibinfo {author} {\bibfnamefont {R.~A.}\ \bibnamefont {Foty}},
  \bibinfo {author} {\bibfnamefont {M.~S.}\ \bibnamefont {Steinberg}}, \ and\
  \bibinfo {author} {\bibfnamefont {E.-M.}\ \bibnamefont {Schoetz}},\
  }\href@noop {} {\bibfield  {journal} {\bibinfo  {journal} {Proc. Natl. Acad.
  Sci. U.S.A.}\ }\textbf {\bibinfo {volume} {107}},\ \bibinfo {pages} {12517}
  (\bibinfo {year} {2010})}\BibitemShut {NoStop}%
\bibitem [{\citenamefont {Fletcher}\ \emph {et~al.}(2014)\citenamefont
  {Fletcher}, \citenamefont {Osterfield}, \citenamefont {Baker},\ and\
  \citenamefont {Shvartsman}}]{Vertex_model}%
  \BibitemOpen
  \bibfield  {author} {\bibinfo {author} {\bibfnamefont {A.~G.}\ \bibnamefont
  {Fletcher}}, \bibinfo {author} {\bibfnamefont {M.}~\bibnamefont
  {Osterfield}}, \bibinfo {author} {\bibfnamefont {R.~E.}\ \bibnamefont
  {Baker}}, \ and\ \bibinfo {author} {\bibfnamefont {S.~Y.}\ \bibnamefont
  {Shvartsman}},\ }\href@noop {} {\bibfield  {journal} {\bibinfo  {journal}
  {Biophys. J.}\ }\textbf {\bibinfo {volume} {106}},\ \bibinfo {pages} {2291}
  (\bibinfo {year} {2014})}\BibitemShut {NoStop}%
\bibitem [{\citenamefont {Giavazzi}\ \emph {et~al.}(2018)\citenamefont
  {Giavazzi}, \citenamefont {Paoluzzi}, \citenamefont {Macchi}, \citenamefont
  {Bi}, \citenamefont {Scita}, \citenamefont {Manning}, \citenamefont
  {Cerbino},\ and\ \citenamefont {Marchetti}}]{giavazzi2018flocking}%
  \BibitemOpen
  \bibfield  {author} {\bibinfo {author} {\bibfnamefont {F.}~\bibnamefont
  {Giavazzi}}, \bibinfo {author} {\bibfnamefont {M.}~\bibnamefont {Paoluzzi}},
  \bibinfo {author} {\bibfnamefont {M.}~\bibnamefont {Macchi}}, \bibinfo
  {author} {\bibfnamefont {D.}~\bibnamefont {Bi}}, \bibinfo {author}
  {\bibfnamefont {G.}~\bibnamefont {Scita}}, \bibinfo {author} {\bibfnamefont
  {M.~L.}\ \bibnamefont {Manning}}, \bibinfo {author} {\bibfnamefont
  {R.}~\bibnamefont {Cerbino}}, \ and\ \bibinfo {author} {\bibfnamefont
  {M.~C.}\ \bibnamefont {Marchetti}},\ }\href@noop {} {\bibfield  {journal}
  {\bibinfo  {journal} {Soft matter}\ }\textbf {\bibinfo {volume} {14}},\
  \bibinfo {pages} {3471} (\bibinfo {year} {2018})}\BibitemShut {NoStop}%
\bibitem [{\citenamefont {Li}\ and\ \citenamefont
  {Ciamarra}(2018)}]{Ourwork_PRM}%
  \BibitemOpen
  \bibfield  {author} {\bibinfo {author} {\bibfnamefont {Y.-W.}\ \bibnamefont
  {Li}}\ and\ \bibinfo {author} {\bibfnamefont {M.~P.}\ \bibnamefont
  {Ciamarra}},\ }\href@noop {} {\bibfield  {journal} {\bibinfo  {journal}
  {Phys. Rev. Mater.}\ }\textbf {\bibinfo {volume} {2}},\ \bibinfo {pages}
  {045602} (\bibinfo {year} {2018})}\BibitemShut {NoStop}%
\bibitem [{\citenamefont {Allen}(1987)}]{Allen_book}%
  \BibitemOpen
  \bibfield  {author} {\bibinfo {author} {\bibfnamefont {M.}~\bibnamefont
  {Allen}},\ }\href@noop {} {\emph {\bibinfo {title} {Computer Simulation of
  Liquids}}}\ (\bibinfo  {publisher} {Oxford University Press, Oxford},\
  \bibinfo {year} {1987})\BibitemShut {NoStop}%
\bibitem [{\citenamefont {Schling}(2011)}]{boost}%
  \BibitemOpen
  \bibfield  {author} {\bibinfo {author} {\bibfnamefont {B.}~\bibnamefont
  {Schling}},\ }\href@noop {} {\emph {\bibinfo {title} {The Boost C++
  Libraries}}}\ (\bibinfo  {publisher} {XML Press},\ \bibinfo {year}
  {2011})\BibitemShut {NoStop}%
\bibitem [{sm()}]{sm}%
  \BibitemOpen
  \href@noop {} {}\bibinfo {note} {See Supplemental Material at http://... for
  additional information about bond-orientational correlation functions and
  $\tau_p$ denpendence of the supercooled dynamics.}\BibitemShut {Stop}%
\bibitem [{\citenamefont {Mermin}\ and\ \citenamefont {Wagner}(1966)}]{MWPrl}%
  \BibitemOpen
  \bibfield  {author} {\bibinfo {author} {\bibfnamefont {N.~D.}\ \bibnamefont
  {Mermin}}\ and\ \bibinfo {author} {\bibfnamefont {H.}~\bibnamefont
  {Wagner}},\ }\href@noop {} {\bibfield  {journal} {\bibinfo  {journal} {Phys.
  Rev. Lett.}\ }\textbf {\bibinfo {volume} {17}},\ \bibinfo {pages} {1133}
  (\bibinfo {year} {1966})}\BibitemShut {NoStop}%
\bibitem [{\citenamefont {Shiba}\ \emph {et~al.}(2016)\citenamefont {Shiba},
  \citenamefont {Yamada}, \citenamefont {Kawasaki},\ and\ \citenamefont
  {Kim}}]{Shiba}%
  \BibitemOpen
  \bibfield  {author} {\bibinfo {author} {\bibfnamefont {H.}~\bibnamefont
  {Shiba}}, \bibinfo {author} {\bibfnamefont {Y.}~\bibnamefont {Yamada}},
  \bibinfo {author} {\bibfnamefont {T.}~\bibnamefont {Kawasaki}}, \ and\
  \bibinfo {author} {\bibfnamefont {K.}~\bibnamefont {Kim}},\ }\href@noop {}
  {\bibfield  {journal} {\bibinfo  {journal} {Phys. Rev. Lett.}\ }\textbf
  {\bibinfo {volume} {117}},\ \bibinfo {pages} {245701} (\bibinfo {year}
  {2016})}\BibitemShut {NoStop}%
\bibitem [{\citenamefont {Vivek}\ \emph {et~al.}(2017)\citenamefont {Vivek},
  \citenamefont {Kelleher}, \citenamefont {Chaikin},\ and\ \citenamefont
  {Weeks}}]{Weeks_longwave}%
  \BibitemOpen
  \bibfield  {author} {\bibinfo {author} {\bibfnamefont {S.}~\bibnamefont
  {Vivek}}, \bibinfo {author} {\bibfnamefont {C.~P.}\ \bibnamefont {Kelleher}},
  \bibinfo {author} {\bibfnamefont {P.~M.}\ \bibnamefont {Chaikin}}, \ and\
  \bibinfo {author} {\bibfnamefont {E.~R.}\ \bibnamefont {Weeks}},\ }\href@noop
  {} {\bibfield  {journal} {\bibinfo  {journal} {Proc. Natl. Acad. Sci. U. S.
  A.}\ }\textbf {\bibinfo {volume} {114}},\ \bibinfo {pages} {1850} (\bibinfo
  {year} {2017})}\BibitemShut {NoStop}%
\bibitem [{\citenamefont {Illing}\ \emph {et~al.}(2017)\citenamefont {Illing},
  \citenamefont {Fritschi}, \citenamefont {Kaiser}, \citenamefont {Klix},
  \citenamefont {Maret},\ and\ \citenamefont {Keim}}]{Keim_MW}%
  \BibitemOpen
  \bibfield  {author} {\bibinfo {author} {\bibfnamefont {B.}~\bibnamefont
  {Illing}}, \bibinfo {author} {\bibfnamefont {S.}~\bibnamefont {Fritschi}},
  \bibinfo {author} {\bibfnamefont {H.}~\bibnamefont {Kaiser}}, \bibinfo
  {author} {\bibfnamefont {C.~L.}\ \bibnamefont {Klix}}, \bibinfo {author}
  {\bibfnamefont {G.}~\bibnamefont {Maret}}, \ and\ \bibinfo {author}
  {\bibfnamefont {P.}~\bibnamefont {Keim}},\ }\href@noop {} {\bibfield
  {journal} {\bibinfo  {journal} {Proc. Natl. Acad. Sci. U. S. A.}\ }\textbf
  {\bibinfo {volume} {114}},\ \bibinfo {pages} {1856} (\bibinfo {year}
  {2017})}\BibitemShut {NoStop}%
\bibitem [{\citenamefont {Li}\ \emph {et~al.}(2019)\citenamefont {Li},
  \citenamefont {Mishra}, \citenamefont {Sun}, \citenamefont {Zhao},
  \citenamefont {Mason}, \citenamefont {Ganapathy},\ and\ \citenamefont
  {Pica~Ciamarra}}]{ywli_PNAS}%
  \BibitemOpen
  \bibfield  {author} {\bibinfo {author} {\bibfnamefont {Y.-W.}\ \bibnamefont
  {Li}}, \bibinfo {author} {\bibfnamefont {C.~K.}\ \bibnamefont {Mishra}},
  \bibinfo {author} {\bibfnamefont {Z.-Y.}\ \bibnamefont {Sun}}, \bibinfo
  {author} {\bibfnamefont {K.}~\bibnamefont {Zhao}}, \bibinfo {author}
  {\bibfnamefont {T.~G.}\ \bibnamefont {Mason}}, \bibinfo {author}
  {\bibfnamefont {R.}~\bibnamefont {Ganapathy}}, \ and\ \bibinfo {author}
  {\bibfnamefont {M.}~\bibnamefont {Pica~Ciamarra}},\ }\href@noop {} {\bibfield
   {journal} {\bibinfo  {journal} {Proc. Natl. Acad. Sci. U. S. A.}\ }\textbf
  {\bibinfo {volume} {116}},\ \bibinfo {pages} {22977} (\bibinfo {year}
  {2019})}\BibitemShut {NoStop}%
\bibitem [{\citenamefont {Ediger}(2000)}]{Ediger}%
  \BibitemOpen
  \bibfield  {author} {\bibinfo {author} {\bibfnamefont {M.~D.}\ \bibnamefont
  {Ediger}},\ }\href@noop {} {\bibfield  {journal} {\bibinfo  {journal} {Annu.
  Rev. Phys. Chem.}\ }\textbf {\bibinfo {volume} {51}},\ \bibinfo {pages} {99}
  (\bibinfo {year} {2000})}\BibitemShut {NoStop}%
\end{thebibliography}
